# Role of Free and Open Source Software in Computer and Internet Security

Muhammad Farooq-i-Azam

Department of Computer Science
COMSATS Institute of Information Technology
Lahore, Pakistan.

*fazam@ciitlahore.edu.pk*

**Abstract**

There is no denying the fact that with the widespread usage of computers and the Internet in our daily lives, security of information and data has gained increased attention. Information stored in electronic form is more susceptible to being leaked to unauthorized individuals intentionally or without intent. One of the major reasons for this breach of security has been attributed to proprietary software whose source is available only to the company which made it. Thus you have no surety that the proprietary and pre-compiled software has no hole to help an individual break security of your computer or network. Philosophy of free and open source software as against this provides everyone an opportunity to view the source code for any possible vulnerabilities and compile and change it according to ones need. This paper discusses this philosophy in length with some examples and also some open source tools that help maintain computer and Internet security.

**Keywords:** Open source software and security, open source philosophy of security, secure software.

## 1. Introduction

Security of an information system depends upon its design and the components used for building it. Apart from the hardware, the major components i.e. brain of a computer or digital system is software. Therefore, how this software is written is a major deciding factor in determining the security of a digital system, be it a piece of code for some ROM, an operating system for a network device like a router or just an application like a web browser.

In 1996, the enquiry board, which reviewed the failed maiden flight of the Ariane 5 launcher of Euproean Space Agency, recommended that the definition of critical components should include software [14].

What makes a software secure or vulnerable depends upon two things:
(a) How the software is designed.
(b) The way it is implemented.
One example of software vulnerabilities is buffer overflows, which may be introduced in the software because of inefficient and poor programming or merely because of a function call which is inherently insecure e.g. strcpy().

Our goal here is to discuss the open source philosophy of security i.e. making the source code public for review by everybody versus the closed software strategy which conceals the source code for no one to see and review.

## 2. Security – Is it improtant?

The question needs to be addressed from many perspectives. Some of these that can be immediately categorized in order of increasing importance are: home users, small and medium enterprises, corporate and multinational companies, governments, military, etc.

While considering the role for home users, computers and digital systems like mobile phones, PDAs etc. have changed dramatically the way we live our lives. Most of the information that used to be only on paper or in hard files is now shared on the computers and the Internet. Be it the accounting information of its customers by bank, transactions history for online banking, examination results of its students by a university or college, sensitive records of police, armed forces, etc. almost everything finds its way to a computer file.

Not only this, we use computers and the Internet for email, instant messaging and voice communication. Moreover, land mobile phones are connected to the

Internet. Even small household items like oven, refrigerators, washing machines, etc. are expected to be networked and connected to the Internet in very near future [7]. Sensitive state information, military command and control system, nuclear plants, almost everything in ministries, government departments, etc. is now connected to the Internet.

With this proliferation of computers everywhere, new methods have been invented to break into computers and use the gained information to the intruder's advantage and according to a survey [5], some even think that the Internet is a conspiracy by an alien society to get into our sociological structure.

Means have been developed to hack cell phones which means your conversations can now be easily eavesdropped by an individual sitting anywhere [13][14]. Even car engines which are increasingly using embedded processors and computers can be subjected to remote attacks [1].

While computer and internet security might not be as important from the perspective of a home user, it is of tremendous value for small businesses, corporate and multinational companies, governments, military, etc. where millions and billions of rupees or dollars or even state secrets are at stake. Consider a bank's security system being compromised and money being transferred to some other account in some other bank. Or imagine a scence where a nuclear plants control system is taken over.

In fact, there are already reports of such security breaches. For example, it was widely reported in the US media that the power failure in North America on August 14, 2003 which left almost 50 million people and the industry without electricity was the work of sabotage by the Internet and SCADA hackers [19].

Again, the same year, a nuclear power plant in Ohio, USA was intruded by an internet worm [20].

These are only known and reported incidents. There may be many such episodes which have not been made public. Banks do not publicize their break-ins so as not to damage their reputation. Governments do not discuss these issues in public as as not to demoralize their people.

It was not without purpose that the premier intelligence agency of USA, CIA conducted first Internet war games this year [12].

### 2.1. The three dimesions of security

Security of a computer system can be viewed from three perspectives:

1- Computer hardware: that would include the processor, its instruction set, architecture, secondary processors, ROMs, the peripherial devices, their processors, ROMs, overall hardware design of the system, etc.

2- Computer Software: Operating system and applications.

3- Network protocols and the associated hardware like routers, switches, etc.

Security plays a role at all the three layers. Software used in all these layers plays a paramount importance whether in the ROM, operating system, application level or in the implementation of protocols in routers and operatin systems.

### 3. How to be secure?

All the things related to security described are serious issues and need to be addressed. To make our computer systems more secure we need to make the software that it runs, more reliable and free of errors and bugs.

### 3.1 Security through obscurity

First measure that a beginner and new software engineer would think of is to make the software code secret for no one to see. This is because, this is how traditional security measures are taken. To save your automobile from thieves, you lock it up in garage. Similarly, to secure your important papers you put them under lock and key in a safe or locker.

This paradigm of security is what is commonly known as *security through obscurity*. While such a scheme might work well in the above situations, it does not produce good results in the case of software.

### 3.2 Open source philosophy of security

Though it is a debate, it is now being widely accepted that instead of keeping the code secret, making it available to everyone for review and changes helps in making it more secure. While it might be hard to percieve this at first, consider the case of scientific research. Research produced by scientists is reviewed by peers for any errors and flaws, corrections are made and the same iterative process is followed before it is acknowledged or is applied in the industry. Same is the case with security of the software. If the code is made secret as in the case of proprietry software, either it is not reviewed at all or possibly by only a handful of software enineers and programmers. Even then you do not have a guarantee of a back door being left intentionally. In fact, such an incident was reported in April 2000, when it was discovered that Microsoft programmers have inserted a back door in their FrontPage Web server software [23]. The flaw was discovered four years after the release of software and this long period of time was due to the reason that the software code was proprietry and secret. Had it been open, there would have been no back door in the first place as a sofware code which is available for public review cannot have apparent provision for its own abuse.

A similar approach of openness has since long been applied to cryptographic algorithms. In cryptography, there is a maxim that security of an algorithm should not depend upon its secrecy. Therefore, famous encryption algorithms like RSA, SSL, etc. are not secrets. Everyone knows the algorithm. Only the keys are secret. Hence the test applied to cryptographic algorithms is:
1- Publish the algorithm and the source code.
2- Programmers are encouraged to find vulnerabilities and errors in the algorithm and code.
3- After the algorithm and code has been thoroughly reviewed and it has been shown that it cannot be compromised, only then it is approved.

Open source software goes under the same test as the above for cryptographic programs. The source code of the program is freely available to everyone to find errors and to fix them. As a result, it goes under a tighter scrutiny and as a result bugs and errors if there are some are purged in the process and the resulting code is more secure.

## 4. Open source security advantages

Though there are many, main points and the merits are described below:
1. Consider the case of proprietry software being installed on your computer for which no source code has been provided. As a result, there is no guarantee that there is no hole left in the code intentionally to spy on people or unintentionally as a result of poor design and programming. Indeed, some think that companies like Microsoft and Intel intentionally leave backdoors in their software under the directives of US government so that the activities of computer users can be spied on. If the source code is concealed as above, even then it can be dissembled and reverse-engineered by crackers and attackers to discover possible exploits. But as against this, take the case of open source software. It is quite obvious that nobody would leave a backdoor in the software for everyone to discover. Consequently, there is an implicit guarantee that the code is clean of any intentional backdoors being left in it. Further, any other bugs are also removed as a result of peer process.

2. When a software code is open to be seen and evaluated by everyone, the developer and programmer takes every care to make it nice, clean and secure because reputation of the programmer is at stake. If there are mistakes and errors left in the code, the whole developers community is out there to discover it. There is no such concept with closed programs.
3. If the software code is made secret, only crackers and attackers discover the holes and they would seldome publicize them i.e. only the bad guys discover mistakes and if there are not publicized, there are no fixes for them too. Whereas in the case of open source software, good guys also get a chance to discover the mistakes and then apply fixes to them.
4. Finding and fizing vulnerabilities and errors in popular software is one way for open source programmers to earn the respect of their peers and the community. Therefore, it provides some motivation to examine source code and improvise upon it.
5. The source code of an open source software can be examined for vulnerabilities in one of two cases:
(a) The complete source code of a software is examined for any vulnerabilities. If the software spans millions of lines of code it is a very tedious and time consuming process.
(a) If there is a known vulnerability being exploited, the exploit can be reverse engineered to find the vulnerability. In this the vulnerability is discovered indirectly and fixed. This is an easier case than the previous one.
6. Because it is open to change from everybody, open source software is very diverse in nature. For example, there are many Unix like operating systems, like, Linux, Solaris, OpenBSD, FreeBSD, etc. Further, Linux has many of its own distributions. Therefore, it is not a good choice for crackers and writers of viruses and malware. It is because of this reason that there has seldom been a case of Linux, Unix or other open source software virus of worm as against the case of proprietry Windows for which multitude of viruses are available and keep surfacing every other day.
7. Companies that try to keep their source code secret run into the risk that someone might access the source code, find bugs and exploit them sometimes without letting the company know it. Examples of software being taken away by hackers already exist [24]. CISCO's network was compromised and almost 800 MB of source code was taken away. In a similar episode, source code Miscrosoft Windows was taken away by hackers.

## 5. Are there only benefits?

Though open source philosphy outweighs the secret code paradigm, there are some points that must be taken into account lest we are swayed by our over-zealeousness of advocasy of open source software:
1. Everyone might assume that someone has done the security auditing of the software whereas no one has done. This is specially true of the software which is not easy to understand or software written in languages not much known e.g. Python.
2. The people in the open source community might not be as motivated to examine a particular code and find and fix errors. Whereas there may be many interested to find holes in linux kernel, there may be not enough motivated people to find and fix errors in an obscure editor software. However, please also note that chances of compromise and damages with an obscure and sparingly used softaware are also less.
3. With the benefit of errors and vulnerabilities being fixed, there is possibility that probable attackers would introduce vulnerabilities in the code. Therefore, as a precaution, open source projects normally accept code from those whom they can trust. This trust is usually built from collaboration with each other for some projects over extended period of time. Use of software like CVS and RCS used for version control and to track which programmer made which changes in the code, also helps track this kind of situation.

## 6. Source code scanning tools

The best way to ensure that a software is free of errors and vulnerabilities is to make a manual audit of the code. However, the process may be time consuming and long enough to be impracticle for projects involving length code. There are many automated tools available to scan a piece of code for any possible errors particulalry those which are documented and quite common. Both proprietry and open source tools are available for this purpose. Some of these valuable open source tools are described below [3][17]:

lint is one of the oldest tool which checks inconsisties and errors in the C code. A similar tool called nslint checks errors in DNS files and another tool weblint checks errors in HTML files. A similar source code scanner for C++ code is clint. Pscan and Cqual are similar tools that scan C source code for inconsistencies.

BOON is a tool that can find buffer overflow possibilites in C programs. MOPS finds vulnerabilities in C programs and checks whether a program conforms to paradigm of secure programming.

Flawfinder is a tool built using Python that can be used to audit C and C++ code.

RATS, the Rough Auditing Tool for Security is a source code scanner that can scan C, C++, Perl, PHP and Python source code.

The scanners should be periodically run on the source code during the development life cycle. The scanners will only highlight where the problem lies. Actual rectifaction of the problem still has to be done by the programmer.

## 7. Conclusions

We have discussed the two philosophies of software development with respect to keeping them secure and not prone to exploits. Each of these has its own merits and demerits. However, as we have seen, open source paradigm has far more benefits than the secret code paradigm which may allow a company to leave a hole or backdoor for later exploit. Particularly, for Pakistan, the open source model provides many benefits in the form of free software alongwith transfer of technology i.e. source code for our review and modification which should result in better security for our IT infrastructure.